\documentclass[conference]{IEEEtran}
\IEEEoverridecommandlockouts

\usepackage{url}
\usepackage{orcidlink}

\usepackage[acronym,toc]{glossaries}
\usepackage{colortbl,hhline}
\usepackage{float}
\usepackage{cite}
\newacronym{iot}{IoT}{Internet of Things}
\newacronym{sbc}{SBC}{Single Board Cumputer}
\newacronym{cgar}{CAGR}{Compound Annual Growth Rate}
\newacronym{cnn}{CNN}{Convolutional Neural Network}
\newacronym{dnn}{DNN}{Deep Neural Network}
\newacronym{dl}{DL}{Deep Learning}
\newacronym{ai}{AI}{Artificial Intelligence}
\newacronym{njn}{NJN}{NVIDIA Jetson Nano}
\newacronym{yolo}{YOLO}{You Only Look Once}
\newacronym{jtag}{JTAG}{Joint Test Action Group}
\newacronym{gpio}{GPIO}{General Purpose Input/Output }
\newacronym{arm}{ARM}{Application Response Measurement}
\newacronym{fpga}{FPGA}{Field Programmable Gate Array}
\newacronym{pbp}{PBP}{Performance Benchmark Program}
\newacronym{fps}{fps}{frames per second}
\newacronym{int8}{int8}{8 bit integer}
\newacronym{aws}{AWS}{Amazon Web Services}
\newacronym{cuda}{CUDA}{Compute Unified Device Architecture}
\newacronym{som}{SOM}{System-on-Module}
\newacronym{gpu}{GPU}{Graphics Processor Units}
\newacronym{sdk}{SDK}{Software Developer Kit}
\newacronym{cpu}{CPU}{Central Processor Units}
\newacronym{aiot}{AIoT}{Artificial Intelligence of Things}
\newacronym{mpsoc}{MPSoC}{Multi-Processor system-on-Chip}
\newacronym{acap}{ACAP}{Adaptive Compute Acceleration Platform}
\newacronym{cv}{CV}{Computer Vision}
\newacronym{ml}{ML}{Machine Learning}
\newacronym{pmurl}{PMU-RL}{Power Measurement Utility Reinforcement Learning}
\newacronym{apc}{APC}{Average Power Consumption}
\newacronym{led}{LED}{Light Emitting Diode}
\newacronym{io}{IO}{Input/Output}
\newacronym{lslb}{LSLB}{LynSyn Lite Board}
\makeglossaries

\def\BibTeX{{\rm B\kern-.05em{\sc i\kern-.025em b}\kern-.08em
    T\kern-.1667em\lower.7ex\hbox{E}\kern-.125emX}}
    
\begin{document}

\title{Estimating the Power Consumption of Heterogeneous Devices When Performing AI Inference
\thanks{The authors would like to thank Mr Flemming Christensen and Sundance Multiprocessor Ltd. for providing the LynSyn Lite board, which was crucial for obtaining the results reported in this paper.}}
\author{
\IEEEauthorblockN{Ivica Matic, Francisco de Lemos\orcidlink{0000-0003-1751-764X}, Pedro Machado\orcidlink{0000-0003-1760-3871},\\ Isibor Kennedy Ihianle\orcidlink{0000-0001-7445-8573}, David Ada Adama\orcidlink{0000-0002-2650-857X}, Jordan J. Bird\orcidlink{0000-0002-9858-1231}}
\IEEEauthorblockA{\textit{Department of Computer Science}\\ School of Science and Technology \\Nottingham Trent University\\
Nottingham, UK \\
Emails: ivica.matic2018@my.ntu.ac.uk,\\ \{francisco.lemos,pedro.machado,isibor.ihianle,david.adama,jordan.bird\}@ntu.ac.uk}
}

\maketitle

\begin{abstract}

Modern-day life is driven by electronic devices connected to the internet. The emerging research field of the Internet-of-Things (IoT) has become popular, just as there has been a steady increase in the number of connected devices. Although these devices are utilised to perform \gls*{cv} tasks, it is essential to understand their power consumption against performance. We report the power consumption profile and analysis of the NVIDIA Jetson Nano board while performing object classification. The authors present an extensive analysis regarding power consumption per frame and the output in frames per second using YOLOv5 models. The results show that the YOLOv5n outperforms other YOLOV5 variants in terms of throughput (i.e. 12.34 fps) and low power consumption (i.e. 0.154 mWh/frame).
\end{abstract}

\section{Introduction}
The rise of \gls*{ai} and the continuous generation of Big Data are creating computational challenges. \glspl*{cpu} are not enough to efficiently run state-of-the-art \gls*{ai} algorithms or process all the data generated by a wide range of sensors. World-leading processing technology companies (such as NVIDIA, AMD, Intel, ARM and Xilinx) have been looking closely into the new requirements. They have been pushing the boundaries of technology for delivering efficient and flexible processing solutions. 

Heterogeneous computing refers to the use of different types of processor systems in a given scientific computing challenge. Heterogeneous platforms are composed of different types of computational units and technologies. Such media can be composed of multi-core \glspl*{cpu}, \glspl*{gpu} and \glspl*{fpga} acting as computational units and offering flexibility and adaptability demanded by a wide range of application domains \cite{SicaDeAndrade2018}. These computational units can significantly increase the overall system performance and reduce power consumption by parallelising concurrent operations that require substantial \gls*{cpu} resources over long periods. 

Accelerators like \glspl*{gpu} and \glspl*{fpga} are massive parallel processing systems that enable accelerating portions of code that are parallelisable. Combining \glspl*{cpu} with \glspl*{gpu} and \glspl*{fpga} helps improve the performance by assigning different computational tasks to specialised processing systems. \glspl*{gpu} are optimised to perform matrix multiplications in parallel, which is the major bottleneck in video processing and computer graphics. Normally, images are captured by cameras and the image stream is sent back to the cloud for processing using \gls*{ai} algorithms. This process consists of visual data to a remote endpoint on the internet and obtaining the response, in which inference results are included. Generally, due to the initial hardware cost of obtaining the appropriate equipment and associated energy costs of running inference equipment, it is cost-effective to employ cloud-based \gls*{cv} services than hosting them locally, provided there is internet access. Moreover, many cloud providers, including \gls*{aws}, Azure, and Google Cloud offer free tier \gls*{cv} services, lowering the initial costs, while allowing users to run their experiments and optimise their applications. 

For some small-scale projects, free tier service might be sufficient. The goal of this work is to improve the quality of service for the existing and new embedded devices that require the use of embedded \gls*{cv} capabilities, which are normally provided by the cloud. Bridging the gap between edge deployed \gls*{iot} devices and their cloud-based processing endpoint, has the following advantages: 1) throughput maximisation due to lower latency. Since the processing endpoint is placed much closer to the execution device, the latency is significantly lower; 2) providing \gls*{cv} capabilities in places with no internet access. Since the endpoint devices are internet independent; and 3) reduces the power consumption and data storage needs because the data is processed locally and the volume of post-processed data to be transmitted to the cloud is substantially less than raw data that is streamed directly from cameras.

The \gls*{njn}\footnote{Available online, \protect\url{https://developer.nvidia.com/embedded/jetson-nano-developer-kit}, last accessed 10/09/2023} was selected over other heterogeneous \glspl*{sbc} because it was available in the laboratory where the experiences were performed. The research aimed at evaluating the \gls*{njn}  in terms of power consumption and \gls*{cv} throughput. The remainder of this paper includes the literature review in Section \ref{sec:lr} review, the description of the \gls*{cv} hardware and software used in this work in  Section \ref{sec:cvhs}, the description of the measurement procedure in Section \ref{sec:mp}, the presentation of the results obtained and their evaluation in Section \ref{sec:re}, and finally the conclusions that and future work in Section \ref{sec:cfw}

\section{Literature review}\label{sec:lr}
The number of internet-connected devices being used for customised applications worldwide is rapidly increasing as technology becomes more accessible and simple to use. Although the estimate of 50 billion internet-connected devices by 2020 was contested back in 2016, it must be acknowledged that 46 billion linked devices by the end of 2021 represent a remarkably sizable quantity \cite{nordrum2016}.

The first generation \gls*{iot} devices can only send and collect data for analysis for further processing. Advances in the later generations of \gls*{iot} are remarkable considering the heterogeneous \glspl*{mpsoc} \cite{wolf2009} and \glspl*{acap} \cite{gaide2019} which enable users to perform more complex operations such as running state-of-the-art\gls*{ai} algorithms at the edge \cite{hassan2018}.

The integration of \gls*{ai} with \gls*{iot}, also known as \gls*{aiot}, can have several advantages in various domains and form ubiquitous ecosystems of intelligent devices working together. Nevertheless, there are significant challenges that must be overcome before the full realisation of its potential. \glspl*{aiot} help to improve our lives in a variety of ways, from making daily jobs simpler to enhancing our health and happiness \cite{ghosh2018artificial}. Efforts are being made to develop model compression and acceleration approaches to deploy \gls*{dl} algorithms on mobile and embedded devices to better satisfy the real-time application constraints and user privacy protection \cite{lane2017squeezing,deng2019deep,chen2020deep}. 

\gls*{iot} networks could experience failure as a result of increased data traffic brought on by more \gls*{iot} nodes \cite{nasif2021}. The biggest problem with \gls*{iot} networks is that they might not have enough memory to manage all the transaction data they need to handle. The solution to this challenge is data compression, a process that reduces the number of bits required to represent data \cite{han1510compressing}. Data compression can reduce network bandwidth requirements, increase speed file transfers, and conserve space on storage systems \cite{liu2018efficient}. With a minor accuracy loss, model compression reduces the complexity and resources required to compress those models. Model compression techniques include parameter reduction, encoding, encryption, and quantisation \cite{qiu2016going,zhao2017accelerating}.

\gls*{yolo} \cite{shafiee2017} is a state-of-the-art \gls*{cnn} that accurately detects objects in real-time. This method processes the entire image using a single neural network, then divides it into parts and forecasts bounding boxes and probabilities for each component. The predicted probability weighs these bounding boxes. The technique "only looks once" at the image, since it only does one forward propagation loop through the neural network before making predictions.

The classified objects are displayed after non-max suppression to ensure that each object is only identified once. Yang et al. \cite{Yang2020} propose a method based on YOLOv5 \cite{glenn2022} to recognise faces wearing masks. When people entered a store, they had to stand in front of a camera, and if recognition succeeded, they could enter through the gate. Additionally, the network was able to classify even when a mask is worn but does not cover the nose.

In general, computer tasks consume variable amounts of energy, and the more processing, the more energy is consumed. Yu et al. \cite{Yu2020} proposes the use of the \gls*{lslb} for measuring power consumption and the \gls*{pmurl} algorithm to dynamically adjust the resource utilisation of heterogeneous platforms in order to minimise power consumption. The \gls*{pmurl} algorithm learns from the power consumption patterns and measurements and controls the programmable logic clock states. Each estimated state of the clock is rewarded when the power consumption is decreased without deteriorating the application performance.

The revised literature shows that it is important to better understand the impact of running state-of-the-art \gls*{ai} algorithms at the edge. Therefore, the authors benchmarked five variants of YOLOv5 on the \gls*{njn} and the \gls*{lslb} for measuring power consumption for each YOLOv5 variant and their outputs in \gls*{fps}.

\section{Methodology}\label{sec:cvhs}
The utilised methods and hardware platform are discussed in this section. 

\subsection{Hardware Platform}
The \gls*{njn} is a heterogeneous platform which was designed to run efficient state-of-the-art \gls*{ai} applications. The NVIDIA Jetpack \gls*{sdk} features a full set of libraries for GPU-accelerated computing, Linux drivers, and the Ubuntu operating system. The on-chip \gls*{gpu} can be programmed using NVIDIA's \gls*{cuda} for accelerating complex and parallelisable algorithms. \gls*{cuda} cores are Nvidia's \gls*{gpu} equivalents of \gls*{cpu} cores. They are built to handle several calculations at once, which is an important feature for accelerating \gls*{dl} algorithms.

Nowadays, \gls*{ai} frameworks such as PyTorch and TensorFlow are already integrated with the high-performance NVIDIA libraries to abstract the \gls*{ai} developers from the \gls*{gpu} hardware complexity. Therefore, the \gls*{njn} was selected over other hardware platforms to benchmark the different YOLOv5 architectures. The \gls*{lslb} was connected to the \gls*{njn} power rails for measuring the power consumption during the \gls*{ai} inference. Finally, the \gls*{lslb} profiling tool and the onboard power sensors were used to measure the power consumption whilst running each of the YOLOv5 variants.

\subsection{Deep Learning Architecture} \label{yolo}
YOLOv5 was selected over other \gls*{dl} methods because it is composed of five variants\footnote{Available online, \url{https://github.com/ultralytics/YOLOv5}, last accessed 01/07/2022}, namely the nano, small, medium, large, and extra-large corresponding to YOLOv5n, YOLOv5s, YOLOv5m, YOLOv5l and YOLOv5x respectively. The YOLOv5 was implemented using the Pytorch framework \footnote{Available online, \protect\url{https://pytorch.org/}, last accessed 01/07/2022} for delivering a user-friendly environment and optimised to leverage from the \gls*{gpu} when required. The different YOLOv5 variants were benchmarked using the COCO2017 dataset and the performance of YOLOv5 was performed using the Ultralytics\footnote{Available online, \protect\url{https://ultralytics.com/}, last accessed 01/07/2022} which is the recommended \gls*{ai} and deployment platform. The COCO2017 pre-trained weights were used because the objective of this work is to estimate the power consumption per neural network variant.

\subsection{LynSyn Lite Board}

\gls*{lslb}\footnote{Available online, \protect\url{https://store.sundance.com/product/lynsyn-lite/}, last accessed 01/07/2022} is a power-profiling tool that was designed for monitoring the power consumption of \glspl*{sbc}. Moreover, the \gls*{lslb} can measure both voltages and current from the power rails in the target device under test and is utilised to extract the power profile for that device. The \gls*{lslb} can be connected to the \gls*{njn} through the \gls*{jtag}, but the \gls*{njn} currently lacks \gls*{jtag} debug capabilities, and therefore the measurement will be done by connecting the \gls*{njn} \gls*{gpio} pins.

Although the \gls*{lslb} also serves the purpose of a \gls*{jtag} programming tool, it was specifically developed for power-profiling \gls*{arm} on Xilinx \gls*{fpga} devices. 

\section{Measurements procedure} \label{sec:mp}

\begin{figure*}[h!]
	\centering
	\includegraphics[scale=0.5]{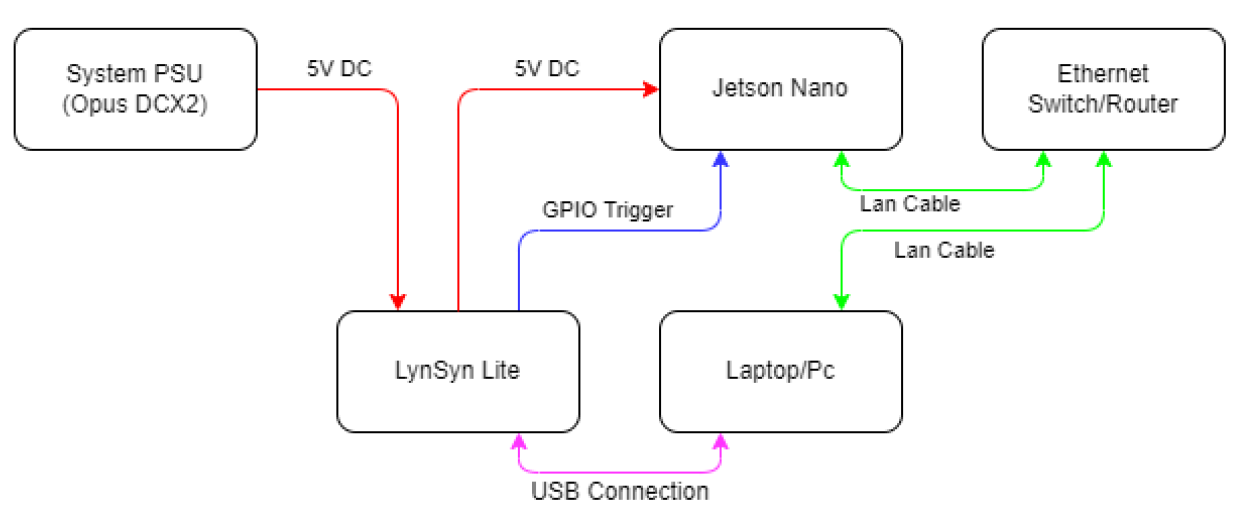}
	\caption{Hardware setup. The power supply is wired to \gls*{njn} through the power rails in the \gls*{lslb}. The \gls*{lslb} is connected to the \gls*{njn} via the \gls*{gpio} headers of both boards. A laptop is used to collect the power measurements from the \gls*{lslb} using the USB port and from \gls*{njn} through the ethernet}
	\label{fig:msh}
\end{figure*}

The \gls*{lslb} viewer application is used to capture the power usage of the \gls*{njn} board running the different variants of the YOLOv5. Figure \ref{fig:msh} shows the hardware setup where it can be seen that the power supply powers \gls*{njn} through the power rails of the \gls*{lslb}. The \gls*{lslb} is also wired to the \gls*{njn} through the \gls*{gpio} of both boards. A laptop is connected to the \gls*{njn} using the ethernet port and to the \gls*{lslb} through the USB bus.

\begin{figure}[h!]
	\centering
	\includegraphics[scale=0.38]{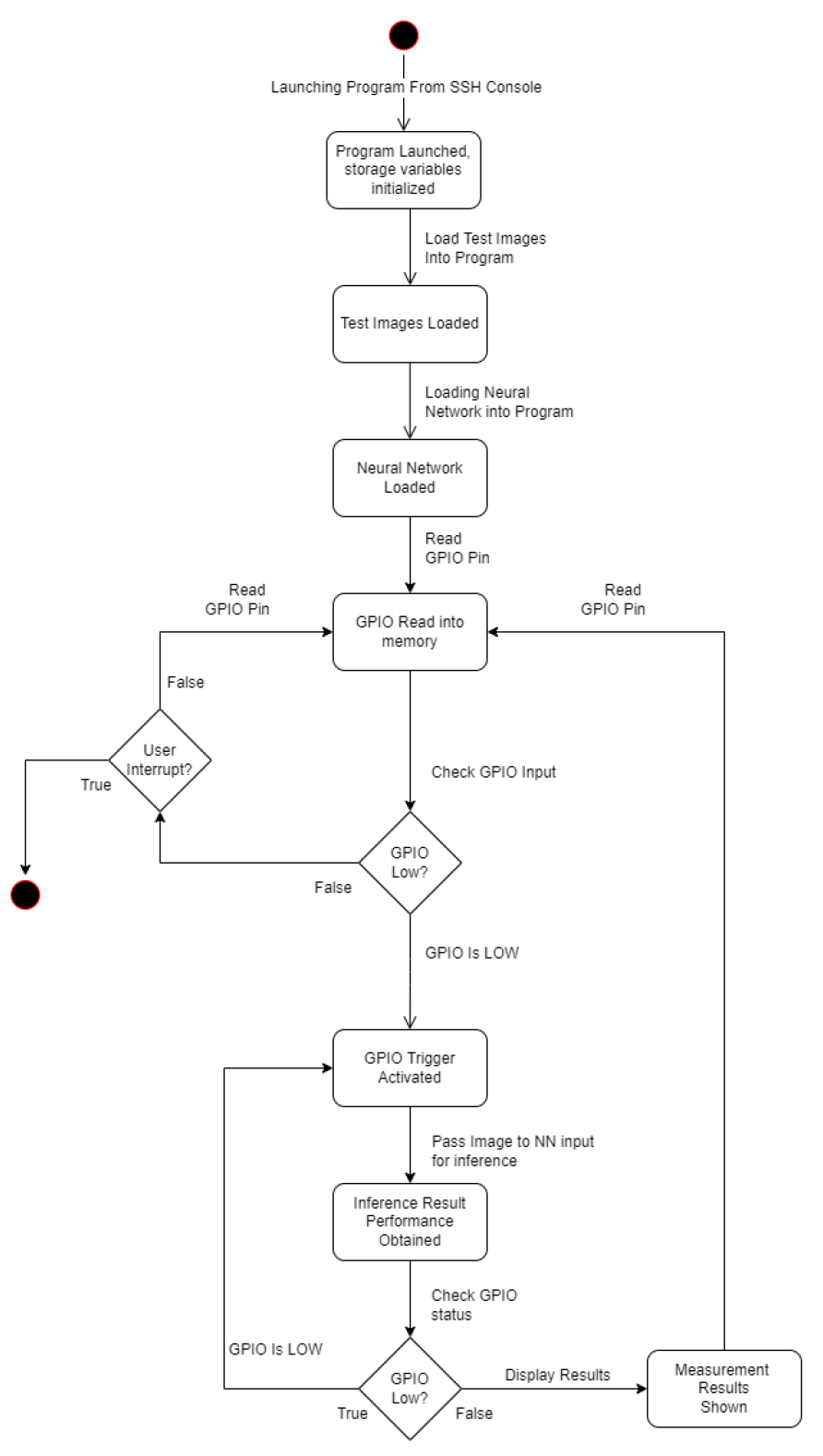}
	\caption{\gls*{pbp} flow diagram.}
	\label{fig:pbp}
\end{figure}

Additionally, a custom \gls*{pbp} application was designed to benchmark the performance of \gls*{njn} and collect the power measurements from the \gls*{lslb} while each of the YOLOv5 variants runs on the \gls*{njn}. Figure~\ref{fig:pbp} depicts a representation of the different states of the \gls*{pbp} algorithm.

The \gls*{pbp} was initialised on the \gls*{njn} using ssh. The \gls*{pbp} loads the COCO2017 test dataset and the target YOLOv5 variants. The test images were exposed to the target YOLOv5 variants while the \gls*{pbp} monitors the signal events exchanged through \gls*{gpio} circuit between the \gls*{njn} and \gls*{lslb}. The \gls*{pbp} starts a thread alongside the target YOLOv5 variants while the \gls*{gpio} is asserted low. The power measures, frame count, and time needed to process each frame are recorded while the signal is asserted low. The performance data is displayed on the screen and the measurements are stored once the process is completed (i.e. once the signal is asserted high). To minimise measurement errors, all measurements were synchronised using \gls*{gpio} and repeated for intervals of 20, 40 and 60 seconds. To provide more data for comparison between the power consumption of different YOLOv5 variants, using the COCO2017 test dataset and a simple $640x480$ video were utilised.

The power consumption was computed using both the \gls*{lslb} and the built-in power sensor data. The relative error $R_{error}$ was computed as function of the instaneous power measure $p_{measured}$ and the average powera for all the measurements $p_{measured}$ using Equation \ref{eq:accuracy} \cite{helfrick1990}:

\begin{eqnarray}
\label{eq:accuracy}
R_{error} [\%] =  \left|1 - \frac{\Delta(p_{meaured} - p_{average})}{p_{average}} \right| . 100
\end{eqnarray}

\section{Results and Evaluations} \label{sec:re}

The power measurement was performed using the procedure described in the previous section using Equation \ref{eq:accuracy}. Multiple measurements were taken per each YOLOv5 variant when running the \gls*{ai} inference on the test images and video. The measurements were carried out in the laboratory with limited precision and therefore is not possible to determine the absolute error. 

\subsection{Performance}

\begin{figure}[]
	\centering
	\includegraphics[scale=0.5]{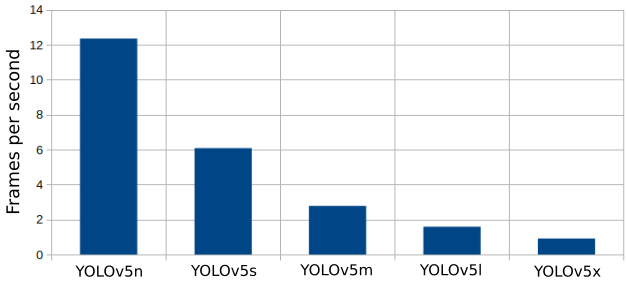}
	\caption{ Performance of various sized YOLOv5 models running on the \gls*{njn}}
	\label{fig:fpsYOLOv5}
\end{figure}

Figure \ref{fig:fpsYOLOv5} shows the that the YOLOv5n has an average throughput of 12 \gls*{fps} when running on the \gls*{njn}.

\begin{figure}[]
	\centering
	\includegraphics[scale=0.5]{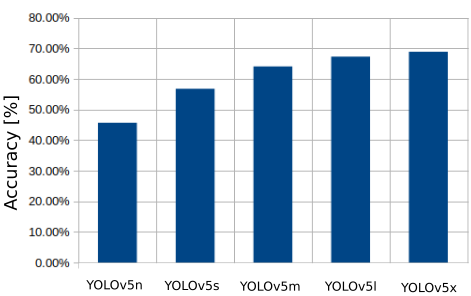}
	\caption{Accuracies of the YOLOv5 variants }
	\label{fig:Y5accuracy}
\end{figure}

The pre-existing trained weights were used as part of the implementation of YOLOv5 variants to determine the accuracies. Figure \ref{fig:Y5accuracy} shows YOLOv5x has the best accuracy while the YOLOv5n has the worst accuracy when tested against the COCO2017 dataset.

\begin{figure}[]
	\centering
	\includegraphics[scale=0.4]{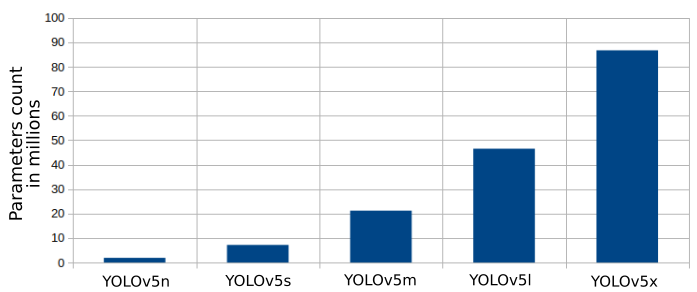}
	\caption{Number of parameters per YOLOv5 variants}
	\label{fig:Y5parameters}
\end{figure}

The average inference speed is 1 \gls*{fps} for the YOLOv5x. The performance of YOLOv5x on \gls*{njn} is substantially lower than the other 4 variants. Nevertheless, the YOLOv5x achieves an accuracy of 70\% when tested against the COCO2007 dataset. The goal of this work is to measure the power consumption and not the issues related to the accuracy of the YOLOv5 variants. YOLOv5x is a suitable variant to use when considering high detection accuracies without the need for real-time performance. The number of parametes per YOLOv5 varant is depicted in Figure \ref{fig:Y5parameters}.

\subsection{Jetson Nano in \gls*{cpu} Only Mode}

\begin{figure}[]
	\centering
	\includegraphics[scale=0.5]{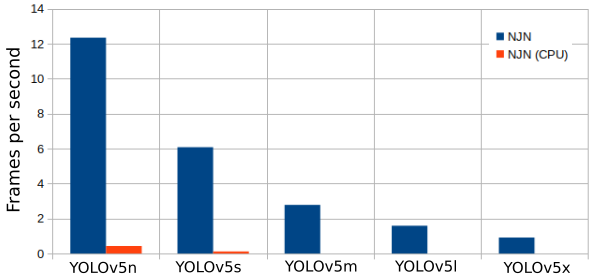}
	\caption{YOLOv5 Inference Performance Comparison between \gls*{cpu} and \gls*{gpu} mode on Jetson Nano}
	\label{fig:21-yolov5}
\end{figure}

The \gls*{njn} features a Quad-core ARM Cortex-A57 processor onboard, so setting the YOLOv5 network to run inference on the \gls*{cpu} allows the acquisition of the performance expected. Figure \ref{fig:21-yolov5} depicts the \gls*{ai} inference performance and that combining the \gls*{cpu} and \gls*{gpu} devices (i.e. \gls*{njn} has a throughput of 12 \gls*{fps} for YOLOv5n) outperforming it by a factor of 30 \gls*{cpu} inference - \gls*{njn} (\gls*{cpu}) has a throughput of 0.4 \gls*{fps} for YOLOv5n.

\subsection{Power Consumption}

Milliwatt-hour per frame (mWh/frame) is the measurement unit for power consumption per each processed image frame. This value is calculated according to Equation \ref{eq:power}, where $P_{frame}$ is the consumed power per image frame, $t_{s}$ the sampling period in seconds, $P_{avg}$ the average power during measurement, $n_{frames}$ the number of frames and constant $K_c$ = 3600 . 1000 used to convert from mWs to Wh.
\begin{eqnarray}
\label{eq:power}
 P_{frame} =  \frac{t_{s}  . P_{avg}  }{n_{frames}.K_c}
\end{eqnarray}

\begin{figure}[]
	\centering
	\includegraphics[scale=0.4]{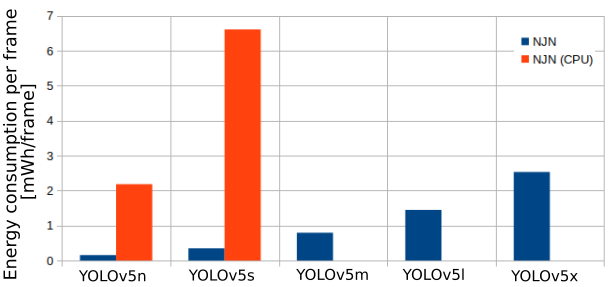}
	\caption{Power consumption per processed frame in mWh/frame}
	\label{fig:23powerperframe}
\end{figure}

Lower mWh/frame values mean less power is required to perform inference per image frame. The results show that YOLOv5x required 2.5 mWh/frame while the smallest YOLOv5n only required 0.15 mWh/frame to perform the same tasks, meaning that YOLOv5n is approximately 16 times more power efficient while running on the \gls*{njn} (see Figure \ref{fig:23powerperframe}).

\begin{figure}[]
	\centering
	\includegraphics[scale=0.4]{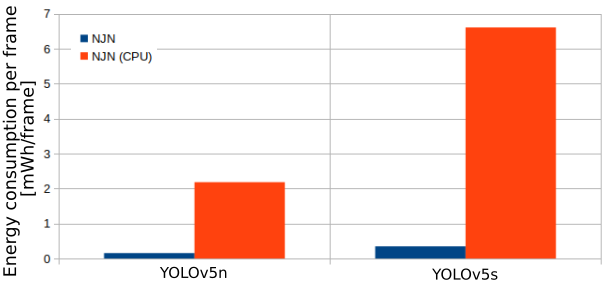}
	\caption{Power consumption per processed frame in mWh/frame (YOLOv5n and YOLOv5s)}
	\label{fig:24}
\end{figure}

The inference using the gls*{cpu} vs. \gls*{cpu} and \gls*{gpu} reduced the power consumption from 2.18 mWh/frame to 0.15 mWh/frame which represents a reduction of 2.03 mWh/frame for the YOLOv5n. This is the benefit of combining heterogeneous platforms at the edge (see Figure \ref{fig:24}).

The power performance is a crucial factor that should be taken into consideration if the system is to be powered by a battery power source. 

The overall results of the power consumption are listed in Table \ref{tbl:6}

\begin{table}[H] \caption{Overall results} \label{tbl:6}
\resizebox{8.9cm}{!}{
\begin{tabular}{|c|c|c|c|c|c|} 
\hline
Model   & Dataset    & Measuring device & Device(s) under test    & \gls*{fps}  & \multicolumn{1}{c|}{\begin{tabular}[c]{@{}c@{}}\gls*{apc}\\  {[}mWh/frame{]}\end{tabular}} \\ \hline
\rowcolor[HTML]{9B9B9B} 
YOLOv5n & COCO2017   & \gls*{lslb}        & CPU + GPU & 11.9 & 159                                                                                 \\ \hline
\rowcolor[HTML]{C0C0C0} 
YOLOv5n & COCO2017   & \gls*{lslb}        & CPU only  & 0.4  & 2183                                                                                \\ \hline
\rowcolor[HTML]{9B9B9B} 
YOLOv5n & COCO2017   & \gls*{njn} Power Sensors   & CPU + GPU & 11.7 & 161                                                                                 \\ \hline
\rowcolor[HTML]{9B9B9B} 
YOLOv5n & Test video &  \gls*{njn} Power Sensors   & CPU + GPU & 13.5 & 143                                                                                 \\ \hline
YOLOv5s & COCO2017   & \gls*{lslb}        & CPU + GPU & 5.9  & 354                                                                                 \\ \hline
\rowcolor[HTML]{C0C0C0} 
YOLOv5s & COCO2017   & \gls*{lslb}        & CPU only  & 0.1  & 6610                                                                                \\ \hline
YOLOv5s & COCO2017   & \gls*{njn} Power Sensors   & CPU + GPU & 5.8  & 363                                                                                 \\ \hline
YOLOv5s & Test video & \gls*{njn} Power Sensors   & CPU + GPU & 6.5  & 328                                                                                 \\ \hline
YOLOv5m & COCO2017   & \gls*{lslb}        & CPU + GPU & 2.7  & 804                                                                                 \\ \hline
YOLOv5m & COCO2017   & \gls*{njn} Power Sensors   & CPU + GPU & 2.7  & 822                                                                                 \\ \hline
YOLOv5m & Test Video & \gls*{njn} Power Sensors   & CPU + GPU & 2.9  & 754                                                                                 \\ \hline
YOLOv5l & COCO2017   & \gls*{lslb}        & CPU + GPU & 1.6  & 1474                                                                                \\ \hline
YOLOv5l & COCO2017   & \gls*{njn} Power Sensors   & CPU + GPU & 1.5  & 1501                                                                                \\ \hline
YOLOv5l & Test Video & \gls*{njn} Power Sensors   & CPU + GPU & 1.7  & 1364                                                                                \\ \hline
YOLOv5x & COCO2017   & \gls*{lslb}        & CPU + GPU & 0.9  & 2562                                                                                \\ \hline
YOLOv5x & COCO2017   & \gls*{njn} Power Sensors   & CPU + GPU & 0.9  & 2580                                                                                \\ \hline
YOLOv5x & Test Video & \gls*{njn} Power Sensors   & CPU + GPU & 1.0  & 2452                                                                                \\ \hline
\end{tabular}}
\end{table}

\subsection{Comparing edge Computing With a Cloud Solution}

Motivated by the low performance of the \gls*{njn} while running bigger YOLOv5 variant networks sizes such as YOLOv5x the Azure \gls*{cv} service was utilised to evaluate the potential performance of the \gls*{njn} sending the frames to the cloud for processing. A batch of measurements used an Azure S1 cloud instance to ensure that there were no bottlenecks in the cloud. The latter is capable of processing 10 frames per second, and the chosen location in the geographically closest region at the time of writing was the UK-SOUTH. The cloud performance is limited mostly by network throughput, achieving only roughly 2 \gls*{fps}. This performance is equivalent to locally running YOLOv5m and YOLOv5l.

Similarly, the power consumption is also equivalent to that of locally running YOLOv5m and YOLOv5l. The \gls*{apc} needed for sending one frame and obtaining the inference results from the cloud is around 2.5 mWh/frame.

\section{Conclusions and Future Work}\label{sec:cfw}
The subject covered in this paper has several impacts on ethical intelligent decision-making. By analysing the power consumption and data throughput of the \gls*{njn} when running different variants of the YOLOv5 model, the authors provide valuable insights into the energy efficiency and performance trade-offs in \gls*{ai} inference at the edge. This information is essential for making informed decisions regarding the deployment of \gls*{ai} systems in resource-constrained environments. Ethical decision-making in this context involves considering the environmental impact and sustainability of \gls*{ai} systems, ensuring they operate efficiently without unnecessarily depleting resources.

Moreover, the study demonstrates that combining both \gls*{cpu} and \gls*{gpu} resources on the \gls*{njn} outperforms using the \gls*{cpu} alone for YOLOv5 inference. This finding highlights the importance of heterogeneity in \gls*{ai} systems for achieving optimal performance. Ethical intelligent decision-making involves considering the most efficient allocation of computational resources to minimising energy consumption and maximising performance while adhering to any constraints or limitations.

From a business perspective, the project opens up future exploitable application areas. The authors show that the \gls*{njn} offers lower power consumption and higher data throughput compared to combining the \gls*{cpu} performance with cloud services like Azure. This finding suggests that the \gls*{njn} can be a cost-effective solution for AI inference at the edge, where real-time processing and low latency are crucial. Businesses can leverage the \gls*{njn} to develop AI applications that require on-device processing, such as autonomous vehicles, surveillance systems, industrial automation, and robotics.

Furthermore, the authors highlight the ease of use and AI acceleration capabilities of the \gls*{njn} and similar NVIDIA boards, thanks to full \gls*{cuda} support. This accessibility and compatibility with popular frameworks and libraries like TensorFlow, Darknet, and PyTorch make it easier for businesses to adopt and integrate AI technologies into their existing workflows. It creates a thriving community with abundant knowledge, tips, and advice, facilitating the development and deployment of AI applications.

For future work, the authors will expand this research to other \gls*{dl} algorithms suitable to be used in edge devices. The authors will also explore the use of \gls*{ml} for estimation of power consumption and reducing power consumption in real-time. Finally, it is also intended to extend this work by performing the same tests using the same methodology on \gls*{mpsoc}, \gls*{acap} and other commercial-off-the-shelf heterogeneous platforms.

\bibliographystyle{IEEEtran}
\bibliography{references}

\end{document}